\documentclass[showpacs,preprintnumbers,amsmath,amssymb]{revtex4}

\usepackage{graphicx}
\usepackage{dcolumn}
\usepackage{bm}

\begin{document}


\title{Diophantine Integrability}
 
\author{R.G. Halburd}
 \email{R.G.Halburd@lboro.ac.uk}
\affiliation{%
Department of Mathematical Sciences,
Loughborough University\\
Loughborough, Leicestershire, LE11 3TU, UK
}%

\date{received January 16, revised March 7, 2005}

\begin{abstract}
The heights of iterates of the discrete Painlev\'e equations over number fields appear to grow no faster than polynomials while the heights of generic solutions of non-integrable discrete equations grow exponentially.
This gives rise to a simple and effective numerical test for the integrability of discrete equations.  
Numerical evidence and theoretical results are presented.  Connections with other tests for integrability and Vojta's dictionary are discussed.
\end{abstract}

\maketitle

Over the past decade and a half several criteria have been suggested as detectors of integrability for maps and discrete equations.  Many of these criteria echo the observation of
Veselov   that 
{\it \dots integrability has an essential correlation with the weak growth of certain characteristics}
\cite{veselov:92}.  A number of authors have studied rational maps and discrete equations for which the degree of the  $n^{\mbox{th}}$ iterate $y_n$ as a rational function of the initial conditions grows no faster than a polynomial in $n$  \cite{growth,veselov:92,hietarintav:98}.  In particular, the algebraic entropy introduced by Hietarinta and Viallet \cite{hietarintav:98} is a measure of this degree growth and is related to Arnold's idea of complexity.

The singularity confinement property of
Grammaticos, Ramani, and Papageorgiou has led to the discovery of many integrable discrete equations \cite{confinement}.  Hietarinta and Viallet \cite{hietarintav:98} have shown that there are non-integrable equations that possess the singularity confinement property, which led them to add the (growth-type) condition of zero algebraic entropy.

 It was suggested in \cite{ablowitzhh:00} that the existence of sufficiently many finite-order meromorphic solutions of a difference equation is a natural analogue of the Painlev\'e property and a detector of integrability.  It has been shown in \cite{halburdk:04} that if an equation of the form
$y(z+1)+y(z-1)=R(z,y(z))$,
where $R$ is rational in its arguments, admits a non-rational finite-order meromorphic solution then either the equation can be transformed to one of the known discrete Painlev\'e equations or $y$ also satisfies a (first-order) discrete Riccati equation. 

Important connections between the differential and discrete Painlev\'e equations, representations of affine Weyl groups, and the geometry of certain rational surfaces have been found in
\cite{noumiy-sakai}.
Costin and Kruskal have suggested that the theory of analyzable functions is the appropriate language in which to describe the Painlev\'e property for discrete equations \cite{costink:02}.
Roberts and Vivaldi have considered maps over finite fields and used orbit statistics to single out detect maps with a polynomial integral of motion \cite{robertsv:03}.

In the present letter a slow-growth property is described which is very easy to test numerically.  It involves considering the iterates of a discrete equation in an appropriate number field (i.e., a finite extension of the rationals) and examining the growth of the height of these iterates.  The height $H(x)$ of an element $x$ of a number field $k$ is a measure of the complexity of $x$.
 
We will, for the most part, only deal with the case $k=\mathbb Q$.  The height of a non-zero rational number $x\in\mathbb Q$ is $H(x)=\max\{|p|,|q|\}$, where $x=p/q$ and $p$ and $q$ have no common factors. The height of 0 is defined to be $H(0)=1$.

\begin{figure}
\includegraphics[width=75mm]{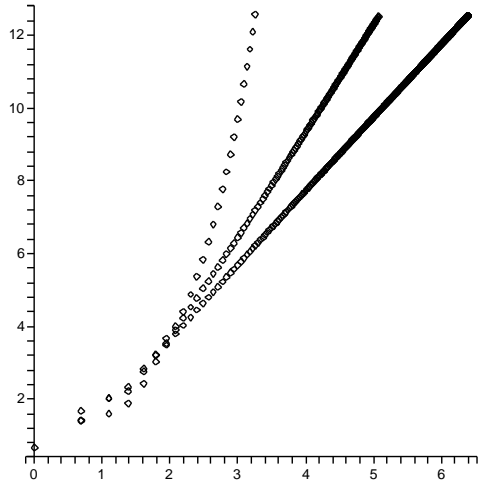}
\caption{\label{fig-dpi}Plot of $\log\log H(y_n)$ vs $\log n$ for equation \eqref{dpi}}
\end{figure}

We begin by considering the growth of heights of iterates $y_n$ of the equation
\begin{equation}
\label{dpi}
y_{n+1}+y_{n-1}=\frac{a_n}{y_n}+b_n,
\end{equation}
where $y_0$ and $y_1$ are given rational numbers and $a_n$ and $b_n$ are chosen to be in ${\mathbb Q}$ for all $n\in {\mathbb Z}$.  This guarantees that all finite iterates $y_n$ are also rational numbers.  In FIG.\,\ref{fig-dpi}, $\log\log H(y_n)$ has been plotted against $\log n$ for three solutions of equation \eqref{dpi}.  In each case, the initial conditions are $y_0=2/5$, $y_1=3/7$ but the choices of $a_n$ and $b_n$ differ.

When $a_n$ and $b_n$ are constants, equation \eqref{dpi} can be solved in terms of elliptic functions.  When $a_n=\lambda n+\mu$ and $b_n=\nu$, for constants $\lambda$, $\mu$ and $\nu$, then equation \eqref{dpi} is an integrable discrete equation related to the first Painlev\'e equation.  In FIG.\,\ref{fig-dpi} there are two integrable cases ($a_n=3$, $b_n=5$ and $a_n=3+n$, $b_n=5$) corresponding to the asymptotically straight line plots, while the third non-integrable case 
($a_n=3$, $b_n=5+n$) corresponds to the asymptotically non-linear curve.

The above example motivates defining a polynomial discrete equation such as \eqref{dpi} to be {\em Diophantine integrable} if the logarithmic height of iterates, $h(y_n)=\log H(y_n)$, grows no faster than a polynomial in $n$.
This idea is certainly related (and possibly equivalent) to the degree growth/algebraic entropy approaches described above but is much quicker to check numerically for a large number of iterates.
For rational maps over $k={\Bbb Q}$, 
Abarenkova, Angl\`es d'Auriac, Boukraa,  Hassani and  Maillard \cite{abarenkovaabhm:99}
used the heights of iterates
as a measure of complexity and to find special values of a parameter for which a particular map was integrable.  Below we will see that for general rational discrete equations we are led to the growth of heights over arbitrary number fields from an analogy with Nevanlinna theory.  We will use this analogy to prove that a large class of discrete equations are not Diophantine integrable.  However, in contrast to the existence of finite-order meromorphic solutions as a detector of integrability for difference equations, it is very simple and quick to use a symbolic computing package to test (numerically) for Diophantine integrability.

To deal with the case in which an iterate becomes infinite it is natural to work in projective space, however, for the purposes of this letter we will only consider finite iterates.  Many autonomous versions of the discrete Painlev\'e equations are known to be solved in terms of elliptic functions.  In fact, these equations are essentially the addition law on the cubic.  For any infinite sequence of rational points on an elliptic curve such that any iterate is the sum (on the cubic) of the previous two, the logarithmic height grows like $n^2$.

\begin{figure}
\includegraphics[width=75mm]{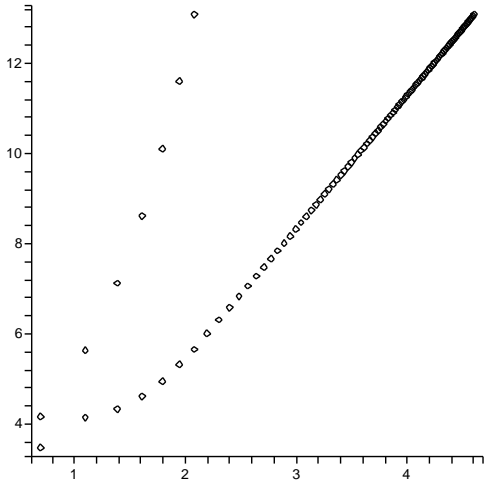}
\caption{\label{fig-qpvi}Plot of $\log\log \max\{H(f_n),H(g_n)\}$ vs $\log n$ for equation \eqref{qpvi}}
\end{figure}

Next we consider the so-called $qP_{V\!I}$ equation, which is the system
\begin{eqnarray}
\label{qpvi}
\frac{f_nf_{n+1}}{cd} &=& \frac{g_{n+1}-\alpha q^{n+1}}{g_{n+1}-\gamma}\, 
\frac{g_{n+1}-\beta q^{n+1}}{g_{n+1}-\delta},\\
\frac{g_ng_{n+1}}{\gamma\delta} &=& \frac{f_n-aq^n}{f_n-c}\,
\frac{f_n-bq^n}{f_n-d},\nonumber
\end{eqnarray}
subject to the constraint 
$q={\alpha\beta\gamma\delta}/{abcd}.$  
The system (\ref{qpvi}) was discovered by  Jimbo and Sakai  as the compatibility condition for an isomonodromy problem \cite{jimbos:96}
and is an integrable discretization of the sixth Painlev\'e equation ($P_{V\!I}$).
FIG.\,\ref{fig-qpvi} is a plot of $\log \log \max\{H(f_n),H(g_n)\}$
against $\log n$ for iterates of equation \eqref{qpvi} with the initial conditions $f_0=2/3$, $g_0=3/4$ and the choice of parameters $(\alpha,\beta,\gamma,\delta,a,b,c,d)=(15/7,4/3,1/2,1,8/7,5/7,2,1/7)$.  The two graphs represent two different choices for $q$, namely, $q=1/2(={\alpha\beta\gamma\delta}/{abcd}$,
i.e., the integrable case corresponding to the asymptotically linear graph) and $q=2$.  Once again we see that the logarithmic heights
$h(f_n)$ and $h(g_n)$ appear to grow polynomially in the integrable case and exponentially in the non-integrable case.

%

Now we will discuss some fundamental identities concerning heights and discrete equations.
Let
$$
R:=
\frac{a_0+a_1x+\cdots+a_px^p}
{b_0+b_1x+\cdots+b_qx^q},
$$
be an irreducible rational function of $x$ of degree $d=\max\{p,q\}$.
Then 
\begin{equation}
\label{height-rational}
C_1 H(x)^d\le H(R) \le C_2 H(x)^d,
\end{equation}
where $C_1$ and $C_2$ are polynomials in the heights of the coefficients $a_i$, $b_j$.
The second inequality in \eqref{height-rational} is straightforward.  In the first inequality
$C_1$ is proportional to the resultant of the denominator and numerator of $R$.
So the logarithmic height $h(\, \cdot\,)=\log H(\, \cdot\,)$ satisfies
\begin{equation}
\label{logheight}
|h(R)-d\,h(x)|\le\log C,
\end{equation}
where $C$ is a polynomial in $H(a_i)$ and $H(b_j)$.

Consider the first-order discrete equation
\begin{equation}
\label{first}
y_{n+1}=
\frac{a_0(n)+a_1(n)y_n+\cdots+a_p(n)y_n^p}
{b_0(n)+b_1(n)y_n+\cdots+b_q(n)y_n^q},
\end{equation}
where the $a_i$'s and $b_j$'s are in ${\mathbb Q}[n]$ and $a_p$ and $b_q$ are both not identically zero.  It follows that for all integers $n$ larger than some $n_0$, the degree of the right side of equation (\ref{first}) as a function of $y_n$ is independent of $n$.  If the right side of equation (\ref{first}) is written in irreducible form for $n>n_0$ then this degree is $d=\max\{p,q\}$.
Taking the logarithmic height of both sides of equation (\ref{first}) and using equation (\ref{logheight}) gives
$$
h(y_{n+1})=d\,h(y_n)+O(\log n).
$$
So if $H(y_n)$ grows faster than any polynomial in $n$ (i.e., if $h(y_n)$ grows faster than $\log n$)
then $h(y_n)$ grows exponentially unless $d\le 1$.  In this case we are left with the (integrable) discrete
Riccati equation $y_{n+1}=\{a_0(n)+a_1(n)y_n\}/\{b_0(n)+b_1(n)y_n\}$, which can be solved via a second-order linear discrete equation.  Hence the demand that solutions grow no faster than polynomials singles out the discrete Riccati equation in the same way that the Painlev\'e property for differential equations singles out the differential Riccati equation.  Note that any periodic orbit of 
equation \eqref{first}, $h(y_n)=O(1)$.

Osgood \cite{osgood} observed that there is an uncanny formal similarity between the basic definitions and theorems of Diophantine approximation and those of Nevanlinna theory.  Independently, Vojta \cite{vojta:87} constructed a dictionary which provides a detailed heuristic for the ``translation'' of concepts and propositions between the two theories.

The fundamental idea of Nevanlinna theory is that much information about a meromorphic function $f$ is obtained by studying a kind of averaged behaviour of $f$ on the disc $D_r:=\{z\,:\, |z|\le r\}$.
The main tool is the Nevanlinna characteristic $T(r,f)$.  For an entire function, $T(r,f)$ behaves like
$\log\max_{|z|=r}|f(z)|$.  In general, $T(r,f)$ is the sum of two terms, one of which is the average of a certain function of $|f|$ on $|z|=r$ and the second is a measure of the number of poles of $f$ in the disc $D_r$.

According to Vojta's dictionary, a statement in Nevanlinna theory about the Nevanlinna characteristic of a meromorphic function corresponds to a statement in Diophantine approximation about an infinite set of numbers in a number field.
In \cite{ablowitzhh:00} it has been argued that a natural analogue of the Painlev\'e property for difference equations is the existence of sufficiently many finite-order meromorphic solutions.  Suppose that, in going from a difference equation for $y(z)$ (in the complex plane) to the corresponding discrete equation for $y_n$ (in which the independent variable is restricted to the integers), we find that all iterates $y_n$ are in some number field $k$ for initial values chosen in $k$.  Via Vojta's dictionary, the statement that $y(z)$ is a finite-order meromorphic function corresponds to the statement that $h(y_n)$
grows no faster than a polynomial.  Hence Diophantine integrability is the natural analogue for discrete equations of the finite-order growth Painlev\'e-type condition for difference equations.

We now derive a number of results about Diophantine integrability by exploiting this formal similarity with Nevanlinna theory.
It is straightforward to check that for any rational numbers $x_1,\ldots,x_N$,
the height satisfies  
$H\left(\sum_{j=1}^N x_j\right)\le N\prod_{j=1}^N H(x_j)$ and
$H\left(\prod_{j=1}^N x_j\right)\le \prod_{j=1}^N H(x_j)$.  It follows that the logarithmic height $h$ has the following properties.
\begin{eqnarray}
h\left(\sum_{j=1}^N x_j\right)&\le& \sum_{j=1}^N h(x_j)+\log N, \label{height-sum}\\
h\left(\prod_{j=1}^N x_j\right)&\le& \sum_{j=1}^N h(x_j).\label{height-prod}
\end{eqnarray}
Note that if we replace the rational numbers $x_j$ and the logarithmic height function $h(\,\cdot\,)$ in 
(\ref{height-sum}) and (\ref{height-prod}) by meromorphic functions $f_j$ and the Nevanlinna characteristic
$T(r,\,\cdot\,)$ respectively, we obtain two standard identities in Nevanlinna theory.
Similarly, it is natural to consider equation \eqref{logheight} to be the Diophantine analogue of the
Valiron-Mohonko Theorem \cite{valironm}.  These are some of the fundamental results used in 
\cite{yanagihara,ablowitzhh:00,heittokangasklrt:01,grammaticostrt:01} to find necessary conditions for the existence of finite-order meromorphic solutions of difference equations.

Consider either of the two second-order discrete equations
\begin{equation}
\label{second}
y_{n+1}+y_{n-1}=R(n,y_n)\mbox{  or  }y_{n+1}y_{n-1}=R(n,y_n),
\end{equation}
where
$R(n,y_n)$ is as in equation \eqref{first}.
Taking the logarithmic height of either of the equations in \eqref{second} and using
equation \eqref{logheight} and the inequality \eqref{height-sum} or \eqref{height-prod} yields
$$
h(y_{n+1})+h(y_{n-1})\ge d\,h(y_n)+O(\log n).
$$
It follows that, provided $h(y_n)$ grows faster than $O(\log n)$, either it grows faster than any polynomial in $n$ or
$d:=\deg_{y_n}(R(n,y_n))\le 2$.  The condition $d\le 2$ is consistent with a number of known integrable discrete Painlev\'e equations such as the special cases of equation \eqref{dpi} described above. 
It is also consistent with the equation
\begin{equation}
\label{dpii}
y_{n+1}+y_{n-1}=\frac{a_n+b_n y_n}{1-y^2_n},
\end{equation}
which is integrable when $a_n$ is a constant and $b_n$ is linear in $n$,
when it is known as the so-called discrete Painlev\'e II equation, $dP_{II}$.

\begin{figure}
\includegraphics[width=75mm]{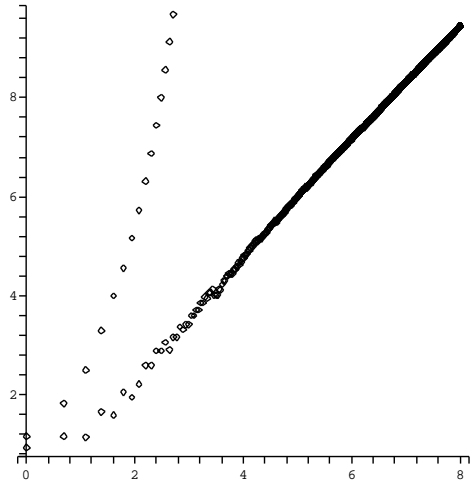}
\caption{\label{fig-dpii}Plot of $\log\log H(y_n)$ vs $\log n$ for equation \eqref{dpii}}
\end{figure}

FIG.\,\ref{fig-dpii} consists of two sequences corresponding to solutions of equation \eqref{dpii} in the non-integrable case $a_n=2n-1$, $b_n=2n^2-2n+3$.  The initial conditions $(y_0,y_1)$ are $(11/12,11/23)$ and $(11/23,11/12)$.
The second initial condition corresponds to the asymptotically linear case in 
FIG.\,\ref{fig-dpii}.
This solution is not generic in that it is also a solution of the first-order (Riccati) discrete equation
$y_{n+1}=(n^2+y_n)/(1-y_n)$.

The conclusion that $d\le 2$ in equations \eqref{second} if $y_n$ has ``slow growth'' (more precisely, $h(y_n)$ grows faster than $O(\log n)$ but is bounded above by a power of $n$) is analogous, both in the conclusion and in the proof, to a result proved in \cite{ablowitzhh:00}.  A number of generalisations of the result in \cite{ablowitzhh:00} have appeared in \cite{heittokangasklrt:01,grammaticostrt:01}.  Many of these results  also have Diophantine analogues, including the results in \cite{grammaticostrt:01} related to $qP_{V\!I}$ (equation \ref{qpvi} above.)

In order to find Diophantine analogues of all the proofs in \cite{grammaticostrt:01}, we need to have one extra identity.  
Fix $N$, let $I$ be the set $I=(1,\ldots,N)$ and for each non-empty $J\subseteq I$ assume $a_J\in{\Bbb Q}$.  Then
$$
H\left(\sum_{J\subseteq I} a_J\prod_{j\in J} x_j\right)\le
|I|
\left(
\prod_{J\subseteq I} H(a_J)
\right)
\left(
\prod_{i=1}^N H(x_i)
\right).
$$
It follows that
\begin{equation}
h\left(\sum_{J\in I} a_J\prod_{j\in J} x_i\right)\le
\sum_{i=1}^N h(x_i)+
\log(C),
\label{height-ramani}
\end{equation}
where $C=\sum_{J\subseteq I}h(a_J)+\log |I|$, which is independent of the $x_i$'s.  Inequality
(\ref{height-ramani}) is a natural Diophantine analogue of a useful inequality in Grammaticos, Tamizhmani, Ramani, and Tamizhmani \cite{grammaticostrt:01} (number 2.11), which is used to obtain the general form of many integrable discrete equations.  The details of this and other calculations will be published elsewhere.

\begin{figure}
\includegraphics[width=75mm]{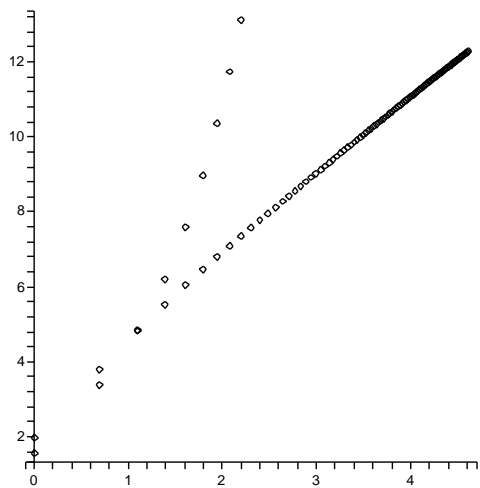}
\caption{\label{fig-dkdv}Plot of $\log\log H(y_{n,n})$ vs $\log n$ for equation \eqref{dkdv}}
\end{figure}

Finally we consider the lattice equation
\begin{equation}
\label{dkdv}
y_{m+1,n+1}=y_{m,n}+\frac 1{y_{m,n+1}}-\frac a{y_{m+1,n}},
\end{equation}
where $a$ is a constant.
 In FIG.\,\ref{fig-dkdv}, $\log\log H(y_{n,n})$ has been plotted against $\log n$ for two different values of $a$ ($a=1$ is the asymptotically linear case, the other is $a=2$) corresponding to the same initial conditions.
The case
$a=1$ is known to be integrable while $a=2$ is not.

\begin{acknowledgments}
This work was supported by the EPSRC and the Leverhulme Trust. 
The author would like to thank the referee for drawing his attention to
\cite{abarenkovaabhm:99}.
\end{acknowledgments}


\begin{thebibliography}{10}
\vskip 3mm

\bibitem{veselov:92}
A.\,P. Veselov,
\newblock {Comm. Math. Phys.} {\bf 145}, 181--193 (1992).

\bibitem{growth}
V.\,I. Arnold,
\newblock{Bol. Soc. Bras. Mat.} {\bf 21},
1--10 (1990);
A.\,P. Veselov,
\newblock {Russian Math. Surveys} {\bf 46}, 1--51 (1991);
G.~Falqui and C.-M. Viallet,
\newblock {Comm. Math. Phys.} {\bf154}, 111--125 (1993);
M.\,P. Bellon and C.-M. Viallet,
\newblock {Comm. Math. Phys.} {\bf 204}, 425--437 (1999);
T. Takenawa.
\newblock{J. Phys. A}{\bf 34}, 10\,533--10\,545 (2001);
Y. Ohta, K.\,M. Tamizhmani, B. Grammaticos, and A. Ramani,
\newblock{Phys. Lett. A} {\bf 262} 152--157 (1999).

\bibitem{hietarintav:98}
J.~Hietarinta and C.~Viallet,
\newblock {Phys. Rev. Lett.} {\bf 81}, 325--328 (1998).

\bibitem{confinement}
B.~Grammaticos, A.~Ramani, and V.~Papageorgiou,
\newblock {Phys. Rev. Lett.} {\bf 67}, 1825--1828 (1991);
A.~Ramani, B.~Grammaticos, and J.~Hietarinta,
\newblock { Phys. Rev. Lett.} {\bf 67}, 1829--1832 (1991).

\bibitem{ablowitzhh:00}
M.\,J. Ablowitz, R.~Halburd, and B.~Herbst,
\newblock {Nonlinearity} {\bf 13}, 889--905 (2000).

\bibitem{halburdk:04}
R.\,G. Halburd and R.\,J. Korhonen,
\newblock{Loughborough University preprint}.

\bibitem{noumiy-sakai}
M. Noumi and Y. Yamada,
\newblock{Commun. Math. Phys.} {\bf 199}, 281--295 (1998);
H. Sakai,
\newblock{Commun. Math. Phys.} {\bf 220}, 165--229 (2001).

\bibitem{costink:02}
O.~Costin and M.~Kruskal,
\newblock {Theoret. and Math. Phys.} {\bf 133}, 1455--1462 (2002).

\bibitem{robertsv:03}
J.\,A.\,G. Roberts and F.~Vivaldi,
\newblock {Phys. Rev. Lett.} {\bf 90}, 034102 (2003).


\bibitem{abarenkovaabhm:99}
N. Abarenkova,  J.-Ch. Angl{\`e}s d'Auriac,  S. Boukraa, 
S. Hassani, and J.-M. Maillard,
 \newblock{Phys. Lett. A}
 {\bf 262}, 44--49 (1999)
 
 
\bibitem{jimbos:96}
M. Jimbo and H. Sakai,
\newblock{Lett. Math. Phys.} {\bf 38}, 145--154 (1996).

\bibitem{osgood}
C.\,F. Osgood,
\newblock{Indian J. Math.}, {\bf 23} 1--15 (1981);
\newblock{J. Number Theory} {\bf 21}, 347--389 (1985).

\bibitem{vojta:87}
P. Vojta,
{\em Diophanitine Approximations and Value Distribution Theory},
\newblock{Lecture Notes in Math.} {\bf 1239}, Springer-Verlag (1987).



\bibitem{valironm}
G. Valiron,
\newblock{Bull. Soc. Math. France} {\bf 59},
17--39 (1931);
A.\,Z. Mohon'ko,
\newblock{Teor. {Funktsi\u\i} Funktsional. Anal. i Prilozhen} {\bf 14},
83--87 (1971).


\bibitem{yanagihara}
N.~Yanagihara,
\newblock {Funkcial. Ekvac.} {\bf 23}, 309--326 (1980);
\newblock {Arch. Rational Mech. Anal.} {\bf 91}, 169--192 (1985).

\bibitem{grammaticostrt:01}
B.~Grammaticos, T.~Tamizhmani, A.~Ramani, and K.\,M. Tamizhmani,
\newblock {J. Phys. A} {\bf 34}, 3811--3821 (2001).

\bibitem{heittokangasklrt:01}
J.~Heittokangas, R.~Korhonen, I.~Laine, J.~Rieppo, and K.~Tohge,
\newblock {Comput. Methods Funct. Theory} {\bf 1}, 27--39 (2001).


\end{thebibliography}

\end{document}